\documentstyle[12pt]{article}

\catcode`\@=11
\long\def\@makefntext#1{
\protect\noindent \hbox to 3.2pt {\hskip-.9pt
$^{{\ninerm\@thefnmark}}$\hfil}#1\hfill}		

 \def\@makefnmark{\hbox to 0pt{$^{\@thefnmark}$\hss}}  

\def\ps@myheadings{\let\@mkboth\@gobbletwo
\def\@oddhead{\hbox{}
\rightmark\hfil\ninerm\thepage}
\def\@oddfoot{}\def\@evenhead{\ninerm\thepage\hfil
\leftmark\hbox{}}\def\@evenfoot{}
\def\sectionmark##1{}\def\subsectionmark##1{}}

\newcounter{sectionc}\newcounter{subsectionc}\newcounter{subsubsectionc}
\renewcommand{\section}[1] {\vspace{0.6cm}\addtocounter{sectionc}{1}
\setcounter{subsectionc}{0}\setcounter{subsubsectionc}{0}\noindent
	{\bf\thesectionc. #1}\par\vspace{0.4cm}}
\renewcommand{\subsection}[1] {\vspace{0.6cm}\addtocounter{subsectionc}{1}
	\setcounter{subsubsectionc}{0}\noindent
	{\it\thesectionc.\thesubsectionc. #1}\par\vspace{0.4cm}}
\renewcommand{\subsubsection}[1]
{\vspace{0.6cm}\addtocounter{subsubsectionc}{1}
	\noindent {\rm\thesectionc.\thesubsectionc.\thesubsubsectionc.
	#1}\par\vspace{0.4cm}}

\newcounter{appendixc}
\newcounter{subappendixc}[appendixc]
\newcounter{subsubappendixc}[subappendixc]

\renewcommand{\appendix}[1] {\vspace{0.6cm}
        \refstepcounter{appendixc}
        \setcounter{figure}{0}
        \setcounter{table}{0}
        \setcounter{equation}{0}
        \renewcommand{\thefigure}{\Alph{appendixc}.\arabic{figure}}
        \renewcommand{\thetable}{\Alph{appendixc}.\arabic{table}}
        \renewcommand{\theappendixc}{\Alph{appendixc}}
        \renewcommand{\theequation}{\Alph{appendixc}.\arabic{equation}}
        \noindent{\bf Appendix \theappendixc #1}\par\vspace{0.4cm}}

\def\abstracts#1{{
	\centering{\begin{minipage}{30pc}\tenrm\baselineskip=12pt\noindent
	\centerline{\tenrm ABSTRACT}\vspace{0.3cm}
	\parindent=0pt #1
	\end{minipage}}\par}}

\renewenvironment{thebibliography}[1]
	{\begin{list}{\arabic{enumi}.}
	{\usecounter{enumi}\setlength{\parsep}{0pt}
\setlength{\leftmargin 1.25cm}{\rightmargin 0pt}
	 \setlength{\itemsep}{0pt} \settowidth
	{\labelwidth}{#1.}\sloppy}}{\end{list}}

\newcounter{itemlistc}
\newcounter{romanlistc}
\newcounter{alphlistc}
\newcounter{arabiclistc}

\newcommand{\fcaption}[1]{
        \refstepcounter{figure}
        \setbox\@tempboxa = \hbox{\tenrm Fig.~\thefigure. #1}
        \ifdim \wd\@tempboxa > 6in
           {\begin{center}
        \parbox{6in}{\tenrm\baselineskip=12pt Fig.~\thefigure. #1}
            \end{center}}
        \else
             {\begin{center}
             {\tenrm Fig.~\thefigure. #1}
              \end{center}}
        \fi}

\newcommand{\tcaption}[1]{
        \refstepcounter{table}
        \setbox\@tempboxa = \hbox{\tenrm Table~\thetable. #1}
        \ifdim \wd\@tempboxa > 6in
           {\begin{center}
        \parbox{6in}{\tenrm\baselineskip=12pt Table~\thetable. #1}
            \end{center}}
        \else
             {\begin{center}
             {\tenrm Table~\thetable. #1}
              \end{center}}
        \fi}

\def\@citex[#1]#2{\if@filesw\immediate\write\@auxout
	{\string\citation{#2}}\fi
\def\@citea{}\@cite{\@for\@citeb:=#2\do
	{\@citea\def\@citea{,}\@ifundefined
	{b@\@citeb}{{\bf ?}\@warning
	{Citation `\@citeb' on page \thepage \space undefined}}
	{\csname b@\@citeb\endcsname}}}{#1}}

\newif\if@cghi
\def\cite{\@cghitrue\@ifnextchar [{\@tempswatrue
	\@citex}{\@tempswafalse\@citex[]}}
\def\citelow{\@cghifalse\@ifnextchar [{\@tempswatrue
	\@citex}{\@tempswafalse\@citex[]}}
\def\@cite#1#2{{$\null^{#1}$\if@tempswa\typeout
	{IJCGA warning: optional citation argument
	ignored: `#2'} \fi}}

\def\fnt#1#2{\footnotetext{\kern-.3em
	{$^{\mbox{\sevenrm #1}}$}{#2}}}

 1
 1
 1

\font\tenbf=cmbx10
\font\tenrm=cmr10
\font\tenit=cmti10

\font\ninerm=cmr9

\def\beq{\begin{equation}}
\def\eeq{\end{equation}}
\def\bl{B - L}
\def\la{~\mbox{\raisebox{-.6ex}{$\stackrel{<}{\sim}$}}~}
\def\ga{~\mbox{\raisebox{-.6ex}{$\stackrel{>}{\sim}$}}~}

\begin{document}
\rightline{UMN-TH-1319/94}
\rightline{November 1994}
\centerline{\tenbf SUPERSYMMETRIC ``SOLUTIONS" TO COSMOLOGICAL PROBLEMS: }
\baselineskip=16pt
\centerline{\tenbf BARYOGENESIS AND DARK MATTER\footnote{To be published in the
proceedings of the Joint US-Polish Workshop on
Physics from Planck Scale to Electroweak Scale,
Warsaw, Poland, September 21-24, 1994, eds. S. Pokorski,
P. Nath, and T. Taylor (World Scientific, Singapore).}}
\vspace{0.8cm}
\centerline{\tenrm KEITH A. OLIVE}
\baselineskip=13pt
\centerline{\tenit School of Physics and Astronomy, University of
Minnesota}
\baselineskip=12pt
\centerline{\tenit Minneapolis, MN 55455, USA}
\vspace{0.9cm}
\abstracts{The possible role of supersymmetry in our understanding
of big bang baryogenesis and cosmological dark matter is explored.
The discussion will be limited to the out-of equilibrium decay scenario in
SUSY GUTs, the decay of scalar condensates, and lepto-baryogenesis as
a means for generating the observed baryon asymmetry.  Attention will
also be focused on neutralino dark matter.
}

\vfil
\rm\baselineskip=14pt
\section{Introduction}
There are several outstanding problems in cosmology which rely
on particle physics solutions.  If supersymmetry (broken as it may be)
is realized in nature, then it is not unreasonable to expect that
supersymmetry plays a non-trivial role in the solutions to these problems.
The two specific problems that I will concentrate upon here are: the
origin of the baryon asymmetry and the nature of dark matter.
The former problem has historically been associated with Grand Unified
Theories (GUTs) and among the original ideas to generate the asymmetry was
the out-of-equilibrium decay scenario\cite{ww}.
 I will  begin, therefore,   with
a look back at the supersymmetric versions of this scenario.
There are also purely supersymmetric solutions to baryogenesis, most
notably is the decay of scalar condensates known as the the Affleck-
Dine (AD) scenario\cite{ad} which
will also be briefly discussed. I will comment on the role of
cosmological inflation
on both the out-of-equilibrium decay and the AD scenarios.  Finally, it is no
longer sufficient
to generate a baryon asymmetry, but one must preserve it in the face
of baryon number violating interactions associated with the standard
electroweak
model\cite{krs}. These interactions, however, open up
new possibilities for generating an asymmetry such as the out-of-equilibrium
decay of superheavy leptons\cite{fy1}.  These possibilities (in the
context of supersymmetry) will also be discussed.

There are many possible solutions to the dark matter problem, many of which
do not involve supersymmetry (nor any new particle physics candidate).
However, the minimal supersymmetric standard model (MSSM) with
unbroken R-parity does offer (in much of the parameter space) a
cosmologically interesting dark matter candidate, the lightest
supersymmetric particle or LSP\cite{ehnos,osi34}. The most likely
choice being the supersymmetric partner of the U(1)-hypercharge
gauge boson, the bino. Though the ``entire" supersymmetric parameter
space will be surveyed, I will focus on the bino as the LSP.
 A curious possibility that the  LSP is a light
photino which is nearly degenerate with the lighter stop quark\cite{fmyy,or2}
 will also be discussed.

\section{Baryogenesis}
\vspace{-0.35cm}
Our best information on the cosmological baryon density comes from
big bang nucleosynthesis. In order to achieve consistency with
the observational determinations of the light element abundances
of deuterium through $^7$Li, the baryon-to photon ratio is restricted to lie
in the range\cite{wssok}
\beq
2.8 \times 10^{-10} < \eta < 4 \times 10^{-10}
\eeq
where $\eta = n_B/n_\gamma$. Combined with the lack of any observed
antimatter (in primary form), our understanding of this small number
is the problem which big bang baryogenesis attempts to solve.

\vspace{-0.35cm}
\subsection{The out-of-equilibrium decay scenario}
\vspace{-0.35cm}
The production of a net baryon asymmetry requires baryon number violating
interactions, C and CP violation and a departure
from thermal equilibrium\cite{sak}.
The first two of these ingredients are contained in GUTs,
the third can be realized in an expanding universe
 where it is not uncommon that interactions
come in and out of equilibrium.
In SU(5), the fact that quarks and leptons are in the same multiplets allows
 for baryon non-conserving interactions such as
$e^{-} + d  \leftrightarrow {\bar u} + {\bar u}$,  etc.,
or decays of the supermassive
 gauge bosons X and Y such as
$ X  \rightarrow e^{-} + d, {\bar u} + {\bar u}$.
 Although today these interactions
are very ineffective because of the very large masses of the X
and Y bosons, in the early Universe when
$T \sim M_{ X} \sim 10^{15}$  GeV these types of interactions
should have been very important.
 C and CP violation is very model dependent.  In the minimal SU(5)
model
the magnitude of C and CP violation is too small to yield a useful value of
$\eta$ and in general the C and CP violation comes
from the interference between
 tree level and one loop corrections.

The departure from equilibrium is very common in the
early Universe when interaction
rates cannot keep up with the expansion rate.  In fact,
the simplest (and most useful)
scenario for baryon production makes use of the fact that a
single decay rate goes out of equilibrium.  It is commonly referred to
 as the out of equilibrium decay scenario\cite{ww}.
The basic idea is that the gauge bosons
 $X$ and $Y$ (or Higgs bosons)
 may have a lifetime long enough to insure that the
inverse decays have already
ceased so that the baryon number is produced by their free decays.

More specifically, let us call $X$, either the gauge
boson or Higgs boson, which produces
the baryon asymmetry through decays.  Let
$\alpha$  be its coupling to fermions.  For $X$ a gauge boson,  $\alpha$
will be the GUT fine structure constant, while for $X$ a Higgs boson,
$(4{\pi \alpha })^{ 1/2}$  will be the Yukawa coupling to fermions.
 If the decay rate for $X$, $\Gamma_{ D}  \simeq   \alpha M_{X}$
is less than the expansion rate of the Universe,
$H \simeq \sqrt{N} T^2/M_P$
(where $N$ is the number of relativistic particles at temperature $T$
and $M_P$ is the Planck mass) at a temperature
$T \sim M_X$
  the decays will occur
 the out-of-equilibrium. Thus the condition  on the superheavy mass is
determined from, $\Gamma {}_{ D} < H$ at $T = M_{ X}$,
or
 \beq
M_{ X} \ga  \alpha M_{ P} (N(M_{ X}))^{ -1/2}
\sim 10^{18} \alpha {\rm GeV}
\label{mxmin}
\eeq
 In this case, we would expect a maximal net baryon asymmetry to be produced
and is given by
\beq
{n_B \over s} \sim {n_B \over N n_\gamma} \sim 10^{-2} \epsilon
\eeq
where $s$ is the entropy density (a better quantity to compare
to in an adiabatically expanding universe) and $\epsilon$ is the baryon
asymmetry produced by an $X,{\bar X}$ decay and represents the degree
of CP violation in the decay.

At least two Higgs five-plets are required to generate sufficient
C and CP violation\cite{nw}. (It is possible within minimal SU(5) to generate a
non-
vanishing $\epsilon$ at 3 loops, however its magnitude would be too small
for the purpose of generating a baryon asymmetry.)  With two five-plets,
$H$ and $H^\prime$, the interference of diagrams of the type in figure 1,
will yield a non-vanishing $\epsilon$,
\beq
\epsilon \propto {\rm Im} ({a^\prime}^\dagger a b^\prime b^\dagger) \ne 0
\eeq
if the couplings $a \ne a^\prime$ and $b \ne b^\prime$.

\vskip 1in

\centerline{{\tenrm Figure 1: One loop contribution to the C and CP
violation with two Higgs five-plets.}}
\vskip .2in

The out-of-equilibrium decay scenario discussed above did not include
the effects of an inflationary epoch.
In the context of inflation\cite{infl},
 one must in addition ensure baryogenesis
after inflation as any asymmetry produced before inflation
would be inflated away along with magnetic monopoles and any other unwanted
relic.
 Reheating after inflation, may require
 a Higgs sector with a relatively light $O(10^{10}-10^{11}) GeV$ Higgs boson.
To see this, consider a simple model in which the inflaton potential
depends on only a single dimensionful parameter $\mu$.
In this case the energy density perturbations produced by inflation
can be roughly estimated to be\cite{cdo}
\beq
{\delta \rho \over \rho} \sim O(100) {\mu^2 \over {M_P}^2}
\label{drho}
\eeq
which when matched to the observed quadrupole moment observed in the
microwave background anisotropy\cite{cobe}
\begin{equation}
{\frac{\delta\rho}{\rho} = (5.4\pm1.6)\times{10^{-6}}}
\end{equation}
fixes the coefficient $\mu$ of the inflaton potential\cite{cdo}:
\begin{equation}
{\frac{\mu^2}{M_P^2} = {\rm few} \times{10^{-8}}}
\label{cobemu}
\end{equation}

       Fixing $({\mu^2}/{M_P^2})$ has immediate general consequences
for inflation\cite{eeno}. For example, the Hubble parameter during inflation,
${{H^2} \simeq (8\pi/3)({\mu^4}/{M_P^2})}$ so that $H \sim
10^{-7}M_P$. The duration of inflation is $\tau \simeq
{M_P^3}/{\mu^4}$, and the number of e-foldings of expansion is $H\tau
\sim 8\pi({M_P^2}/{\mu^2}) \sim 10^{9}$. If the inflaton decay rate
goes as $\Gamma \sim {m_{\eta}^3}/{M_P^2} \sim {\mu^6}/{M_P^5}$, the
universe recovers at a temperature $T_R \sim (\Gamma{M_P})^{1/2} \sim
{\mu^3}/{M_P^2} \sim 10^{-11} {M_P} \sim 10^8 GeV$. Thus,
the light Higgs
is necessary since the inflaton, $\eta$, is typically light ($m_\eta
\sim \mu^2/M_P \sim
O(10^{11})$ GeV),
and the baryon number violating Higgs
would have to be produced during inflaton decay.
Note that a ``light" Higgs is acceptable from the point of view of proton
decay due to its
reduced couplings to fermions.
The out-of-equilibrium decay scenario would now be realized by
Higgs boson decay rather than gauge boson decay and a different sequence
of events. First the inflaton would be required to
decay to Higgs bosons (triplets?) and subsequently
the triplets would decay rapidly by the processes shown in figure 1.
These decays would be well out of equilibrium
as at reheating $T \ll m_H$ and $n_H \sim n_\gamma$\cite{dlnos}.
In this case, the baryon asymmetry is given simply by
\beq
{n_B \over s} \sim \epsilon {n_H \over {T_R}^3}
\sim \epsilon {n_\eta \over {T_R}^3}
\sim \epsilon {T_R \over m_\eta} \sim \epsilon
\left( {m_\eta \over M_P} \right)^{1/2}
\sim \epsilon {\mu \over M_P}\sim 10^{-4} \epsilon
\eeq
where I have substituted for $n_\eta = \rho_\eta/m_\eta
\sim \Gamma^2{M_P}^2/m_\eta$.

In a supersymmetric grand unified SU(5)
 theory, the superpotential $F_Y$ can be expressed in terms
of SU(5) multiplets
\beq
	F_Y  = h_d {\bf H_2 ~{\bar 5}~10}  +  h_u {\bf H_1~10~10}
\eeq
where $10, {\bar 5}, H_1$ and $H_2$ are chiral
supermultiplets for the 10, and ${\bar 5}$ plets of
SU(5) matter fields and the Higgs
5 and ${\bar 5}$ multiplets respectively.
There are now new dimension 5
operators which violate baryon number and lead to proton decay
as shown in figure 2.
The first of these diagrams leads to effective dimension 5 Lagrangian terms
such as
\beq
 {\cal L}_{\rm eff}^{(5)} = {h_u h_d \over M_{H}} ( \tilde q
\tilde q q l)
\eeq
and the resulting dimension 6 operator for proton decay\cite{enr}
\beq
{\cal L}_{\rm eff} = {h_u h_d \over M_{H}} \left( {g_2^2 \over M_{\tilde W}}
 {\rm or} {g_1^2 \over M_{\tilde B}} \right) (  q q q l)
\eeq
As a result of these diagrams the proton decay rate scales as $\Gamma
\sim h^4 g^4/M_H^2 M_{\tilde G}^2$ where $M_H$ is the triplet mass, and
$M_{\tilde G}$ is a typical gaugino mass of order $\la$ 1 TeV.  This rate
however is much too large if $M_H \sim 10^{10}$ GeV.

It is however possible to have a lighter ($O(10^{10}-10^{11})$ GeV)
Higgs triplet needed for baryogenesis in the out-of-equilibrium decay
scenario with inflation.  One needs  two pairs of Higgs five-plets
($H_1, H_2$ and  $H_1^\prime, H_2^\prime$) which
is anyway necessary to have sufficient
C and CP violation in the decays.
By coupling one pair $(H_2$ and $H_1^\prime)$
only to the third generation of fermions
via\cite{nt}
\beq
a {\bf H_1 10 10} + b {\bf H_1^\prime 10_3 10_3} + c {\bf H_2 10_3 {\bar 5}_3}
+ d {\bf H_2^\prime 10 {\bar 5}}
\eeq
proton decay can not be induced by the dimension five operators.
Triplet decay will however generate a baryon asymmetry proportional to
$\epsilon \sim {\rm Im} d c^\dagger b a^\dagger$.

\vskip 1in

\centerline{{\tenrm Figure 2: Dimension 5 and induced dimension 6
graphs violating baryon number.}}
\vskip .2in

\subsection{The Affleck-Dine Mechanism}
\vspace{-0.35cm}
Another mechanism for generating the  cosmological baryon asymmetry
is the decay of scalar condensates as first
proposed by Affleck and Dine\cite{ad}.
This mechanism is truly a product of supersymmetry.
It is straightforward though tedious to show that
  there are many directions in field space such that the scalar potential
 vanishes identically
 when SUSY is unbroken.
 SUSY breaking lifts this degeneracy so that
\begin{equation}
	V  \simeq \tilde{m}^2 \phi^2
\eeq
where $\tilde{m}$ is the SUSY breaking scale and $\phi$ is the direction
 in field space corresponding to the flat direction.
  For large initial values of $\phi$, \ $\phi_o \sim M_{GUT}$,
 a large baryon asymmetry can be generated\cite{ad,l}. This requires
the presence of baryon number violating operators such as $O=qqql$
which are naturally provided for in superymmetric GUTs and such that
$\langle O \rangle \neq 0$.  The decay of these
 condensates through such an operator with an effective quartic coupling
of order ${\tilde m}^2/(\phi_o^2 + M_X^2)$
can lead to a net baryon asymmetry.

The baryon asymmetry produced, is computed by tracking the evolution of the
sfermion condensate, which is determined by
\begin{equation}
\ddot{\phi} + 3H\dot{\phi} = - {\tilde m}^2 \phi
\end{equation}
If it is assumed that the energy density of the Universe is dominated
by $\phi$, then the oscillations will cease, when
\beq
\Gamma_\phi \simeq {{\tilde m}^3 \over \phi^2} \simeq H
\simeq {\rho_\phi^{1/2} \over M_P} \simeq {{\tilde m} \phi \over M_P}
\eeq
or when the amplitude of oscillations has dropped to $\phi_D \simeq (M_P
{\tilde m}^2 )^{1/3}$. Note that the decay rate is suppressed as
fields coupled directly to $\phi$ gain masses $\propto \phi$.
It is now straightforward to compute the baryon to entropy ratio,
\beq
{n_B \over s} = {n_B \over \rho_\phi^{3/4}} \simeq {\lambda \phi_o^2
\phi_D^2 \over {\tilde m}^{5/2} \phi_D^{3/2}} =
{\lambda \phi_o^2 \over {\tilde m}^2} \left({M_P \over \tilde m}\right)^{1/6}
\eeq
and after inserting the quartic coupling,$\lambda$,
\beq
{n_B \over s} \simeq \epsilon {\phi_o^2 \over (M_X^2 + \phi_o^2)}
\left({M_P \over \tilde m}\right)^{1/6}
\eeq
which could be quite large.

In the context of inflation, a couple of significant changes to the scenario
take place. First, it is more likely that the energy density
is dominated by the inflaton rather than the sfermion condensate.
Second, the the initial value (after inflation) of the condensate
$\phi$  can be determined by the inflaton mass $m_\eta$,
${\phi_o}^2 \simeq H^3\tau \simeq m_\eta M_P$.
The baryon asymmetry in the Affleck-Dine
 scenario with inflation becomes\cite{eeno}
\begin{equation}
	\frac{n_B}{s} \sim  \frac{\epsilon {\phi_o}^4 {m_\eta}^{3/2}}
{{M_X}^2 {M_P}^{5/2} \tilde{m}} \sim
 \frac{\epsilon m_\eta^{7/2}}{{M_X}^2 {M_P}^{1/2} \tilde{m}}
  \sim  (10^{-6}-1) \epsilon
\end{equation}
for
 $\tilde{m} \sim (10^{-17}-10^{-16}) M_P$,
 and $M_X \sim (10^{-4}-10^{-3}) M_P$
and $m_\eta \sim (10^{-8} - 10^{-7} ) M_P$.

\subsection{Lepto-baryogenesis}
\vspace{-0.7cm}
\subsubsection{Preservation of the asymmetry}
\vspace{-0.35cm}
The realization\cite{krs} of significant baryon number
violation at high temperature within the
standard model, has opened the door for many new
possibilities for the generation
of a net baryon asymmetry. Electroweak
 baryon number violation
occurs through non-perturbative interactions mediated by ``sphalerons",
 which violate $B + L$ and conserve
$B - L$.  For this reason, any GUT produced asymmetry with $B - L = 0$
may be subsequently erased by sphaleron interactions\cite{am}.

With $B- L = 0$, it is relatively straightforward to see
that the equilibrium conditions
including sphaleron interactions gives zero net baryon number\cite{ht}.
By assigning each particle species a chemical potential, and
using gauge and Higgs
interactions as conditions on these potentials
(with generation indices suppressed),
\begin{eqnarray}
\mu_- + \mu_0 = \mu_W \qquad
\mu_{u_R} - \mu_{u_L} = \mu_0 \qquad
\mu_{d_R} - \mu_{d_L} = -\mu_0 \nonumber \\
\mu_{l_R}-\mu_{l_L} = -\mu_0 \qquad
\mu_{d_L} - \mu_{u_L} = \mu_W \qquad
\mu_{l_L} - \mu_\nu =  \mu_W
\end{eqnarray}
one can write
down a simple set of equations for the baryon and
lepton numbers and electric charge which reduce to:
\begin{eqnarray}
B& = & 12 \mu_{u_L} \nonumber \\
L& = & 3 \mu - 3 \mu_0 \label{mus} \\
Q& = & 6 \mu_{u_L} -2\mu + 14 \mu_0 \nonumber
\end{eqnarray}
where $\mu = \sum \mu_{\nu_i}$.
In (\ref{mus}), the constraint on the weak isospin charge,
$Q_3 \propto \mu_W = 0$ has been employed.
Though the charges $B,L,$ and $Q$ have been written as
chemical potentials, since for small asymmetries,
an asymmetry $(n_f - n_{\bar f})/s \propto \mu_f/T$,
we can regard these quantities as net number densities.

The sphaleron process yields the additional condition,
\beq
 9\mu_{u_L} + \mu = 0
\label{S}
\eeq
which allows one to solve for $L$ and $B-L$ in terms of
$\mu_{u_L}$, ultimately giving
\beq
B = {28 \over 79} \left( B - L \right)
\label{2879}
\eeq
Thus, in the absence of a primordial $B-L$ asymmetry,
the baryon number is erased by equilibrium processes.
Note that barring new interactions (in an extended model)
the quantities ${1 \over 3}B - L_e$, ${1 \over 3}B - L_\mu$,
and ${1 \over 3}B - L_\tau$ remain conserved.

With the possible erasure of the baryon asymmetry when $B-L=0$ in mind,
since minimal SU(5) preserves $B-L$, electroweak
effects require GUTs beyond
SU(5) for the asymmetry generated by the out-of-equilibrium
decay scenario to survive.
GUTs such as SO(10) where a primordial $\bl$ asymmetry
can be generated becomes a promising choice.
The same holds true in the Affleck-Dine mechanism for generating a
baryon asymmetry.
  In larger GUTs there are baryon number violating operators
and associated flat directions\cite{24}.  A specific
example in SO(10) was worked out in detail by Morgan\cite{25}.

Another possibility for preserving a primordial baryon asymmetry
when $\bl = 0$ arises if the asymmetry produced by scalar condensates
in the Affleck-Dine mechanism is large\cite{dmo} ($n_B/s \ga 10^{-2}$).
After the decay of the A-D condensate, the baryon number is
shared among fermion and boson superpartners. However, in equilibrium,
there is a maximum chemical potential $\mu_f = \mu_B = \tilde m$
and  for a large asymmetry, the baryon number density stored in
fermions, $n_{B_f} = {g_f \over 6} \mu_f T^2$ is much less than the total
baryon density. The bulk of the baryon asymmetry is driven into the
$p=0$ bosonic modes  and a Bose-Einstein condensate is formed\cite{dkiri}.
The critical temperature for the formation of this condensate
is given by $n_B \simeq n_{B_b} + n_{B_c} = {g_b \over 3} \tilde{m} T_c^2$
so that,
\beq
n_{B_c} = {g_b \over 3} \left( 1 - \left( {T \over T_c} \right)^2 \right)
 T_c^2
\eeq
At $T < T_c$, most of the baryon number remains in a condensate and
for large $n_B$, the condensate persists down to temperatures
of order 100 GeV. Thus sphaleron interactions are shut off and
a primordial baryon asymmetry is maintained even with $\bl = 0$.
One should note however that additional sources of entropy are required
to bring $\eta$ down to acceptable levels.

\subsubsection{Generating a baryon asymmetry from a primordial lepton
asymmetry}
\vspace{-0.35cm}
Sphaleron interactions also allow
 for new mechanisms to produce a baryon asymmetry.
The simplest of such mechanisms
is based on the decay of a right handed neutrino-like state\cite{fy1}.
This mechanism
is certainly novel in that does not require grand unification at all.
By simply adding to the Lagrangian a Dirac and Majorana mass term
 for a new right handed neutrino state,
\beq
{\cal L} \ni M\nu^c\nu^c + \lambda H L \nu^c
\eeq
the out-of-equilibrium decays $\nu^c \rightarrow L +  H^*$
 and  $\nu^c \rightarrow L^* + H$ will generate a non-zero
lepton number $L \neq 0$. The out-out-equilibrium condition
for these decays translates to $10^{-3} \lambda^2 M_P < M$
and $M$ could be as low as $O(10)$ TeV.
(Note that once again in order to
have a non-vanishing contribution to the C and CP violation
in this process at 1-loop, at least 2 flavors of $\nu^c$ are required.
For the generation of masses of all three neutrino flavors,
3 flavors of $\nu^c$ are required.)
 Sphaleron effects can transfer this lepton asymmetry into a baryon
 asymmetry since now $B - L \neq 0$. A supersymmetric version of
this scenario
has also been described\cite{cdo,mur}.

	The survival of the asymmetry, of course depends on
 whether or not electroweak sphalerons can wash away the asymmetry.
 The persistence of lepton number violating interactions in conjunction
 with electroweak sphaleron effects could wipe out\cite{fy2}
 both the baryon and lepton asymmetry in the
 mechanism described above through effective operators
 of the form $\lambda^2 LLHH/M$.
In terms of chemical potentials, this interaction adds the condition
$\mu_\nu + \mu_0 = 0$.  The constraint comes about by requiring that
this interaction
be out of equilibrium at the time when sphalerons are in equilibrium.
 Otherwise, the additional
condition on the chemical potentials would force the solution $B=L=0$.
To prevent the erasure of the baryon asymmetry, the constraint on
$M/\lambda^2 \ga 3 \times 10^9$ GeV obtained by requiring the $B+L$
violating operators to remain out of equilibrium at least until
right-handed electrons come into equilibrium\cite{clkao23}  leads to a bound
on neutrino masses, $m_\nu \sim \lambda^2 v^2 /M \la 20$ keV, where
$v = 247$ GeV is the Higgs vev.  Similar constraints can be derived
on $R$-parity violating operators\cite{sim}.

In addition to the mechanism described earlier utilizing a right-handed
neutrino decay,
several others are now also available.  In a supersymmetric
extension of the standard
model including a right-handed neutrino, there are numerous
possibilities. Along the
lines of the right-handed neutrino decay, the scalar partner\cite{cdo} or a
condensate\cite{mur} of ${\tilde \nu^c}$'s will easily
generate a lepton asymmetry.
Furthermore if the superpotential contains terms such
 as ${\nu^c}^3 + \nu^c H_1H_2$,
there will be a flat direction violating lepton number\cite{cdo1,cdo}
 \`{a} la Affleck and Dine.
While none of these scenarios require GUTs, those that
involve the out-of equilibrium
decay of either fermions, scalars or condensates must have
the mass scale of the
right-handed neutrino between $10^9$ and about $10^{11}$
GeV, to avoid washing
out the baryon asymmetry later
 and to be produced after
inflation respectively.
In contrast, the decay of the flat direction condensate
(which involves other fields
in addition to ${\tilde \nu}^c$) only works for $10^{11} < M < 10^{15}$ GeV.

\section{Dark Matter}
There are several reasons for postulating the existence of dark matter.
On the theoretical side, if the cosmological density parameter
is one, then the upper bound on the fraction of $\Omega$ in baryons
is restricted by nucleosynthesis to take values\cite{wssok} $\Omega_B < 0.08$
leaving the remainder as non-baryonic dark matter.  Also on the theoretical
side, is the effect of dark matter on the growth of density perturbations.
The problem of making galaxies and clusters is exasperated without dark
matter.
There are also several observational pieces of evidence which include:
galactic rotation curves, $X$-ray emitting hot gas from elliptical
galaxies and clusters, as well as gravitational lensing by dark halos.
What portion of the dark matter is truly non-baryonic is still
unknown, but if in fact $\Omega = 1$, most of the dark matter
would be in the form a new particle candidate.  I will here concentrate
only the supersymmetric candidates. For a more general recent review see:
ref. (31).

Supersymmetric theories introduce several possible candidates.
If $R$-parity (which distinguishes between ``normal" matter and the
supersymmetric partners) is unbroken there is at least one
supersymmetric particle which must be stable.
I will assume R-parity conservation.
The stable
particle (usually called the LSP) is most probably some
linear combination of the only $R=-1$ neutral fermions, the
neutralinos\cite{ehnos}: the wino $\tilde W^3$, the partner of the
 3rd component of the $SU(2)_L$ gauge boson;
 the bino, $\tilde B$, the partner of the $U(1)_Y$ gauge boson;
 and the two neutral Higgsinos  $\tilde H_1$, and $\tilde H_2$.
 Gluinos are expected to be heavier,  $m_{\tilde g} = (\frac{\alpha_3}{\alpha})
 \sin^2 \theta_W M_2$ and do not mix with the other states ($M_2$ is the
soft SUSY breaking $SU(2)$ gaugino mass).
  The sneutrino\cite{snu} is also a possibility but has been excluded as a dark
matter candidate by direct\cite{dir} searches, indirect\cite{indir}
 and accelerator\cite{lep}
searches. For a recent examination of very heavy sneutrino candidates
see ref.(36).

 The the only parameters which determine the
mass and composition of the LSP are; $M_2$, $\mu$ and $\tan \beta$
(assuming the GUT relations among the soft SUSY breaking gaugino masses).
The latter two are the supersymmetric Higgsino mixing mass and the ratio
of the Higgs scalar vevs respectively.
However, for the relic abundance of
LSP's, it is necessary to specify the Higgs (scalar) masses,
and the sfermions masses.
	The LSP can  be expressed as a linear combination
\begin{equation}
		\chi = \alpha \tilde W^3 + \beta \tilde B +
\gamma \tilde H_1 + \delta \tilde H_2
\end{equation}
Pure state LSP possibilities are: The photino\cite{phot}, when $M_2 \rightarrow
0$
\begin{equation}
		\tilde {\gamma} = \tilde W^3\sin\theta_W + \tilde B \cos\theta_W
\end {equation}
and
\beq
	m_{\tilde \gamma} \rightarrow {8 \over 3}
{ {g_1}^2 \over  ( {g_1}^2 + {g_2}^2) } M_2
\eeq
the Higgsino, $\tilde S^0$ \cite{ehnos}, when $\mu  \rightarrow 0$
\begin{equation}
		\tilde S^0 = \tilde H_1 \cos\beta + \tilde H_2 \sin\beta
\end{equation}
and
\beq
	m_{\tilde S}  \rightarrow  {2 v_1 v_2 \over v^2} \mu
\eeq
When $M_2$  is large and $M_2 \ll \mu$
then the bino\cite{osi34},  ${\tilde B}$,
 is the LSP    and
\beq
	m_{\tilde B}  \simeq M_1 	= {5 \over 3}  {\alpha_1 \over \alpha_2}  M_2
\eeq
Finally when $\mu$ is large and $\mu  \ll M_2$
 either the Higgsino state\cite{osi34}
\beq
	{\tilde H}_{(12)}  =  {1 \over \sqrt{2}}  ( {{\tilde H}_1}^0 +
{{\tilde H}_2}^0)  \qquad
 \mu < 0
\eeq
or the state
\beq
{\tilde H}_{[12]}  =  {1 \over \sqrt{2}}  ( {{\tilde H}_1}^0 -
{{\tilde H}_2}^0)  \qquad
 \mu > 0
\eeq
is the LSP depending on the sign of $\epsilon$ and
\beq
m_{\tilde H} \simeq |\mu|
\eeq

\vskip 1in
\centerline{{\tenrm Figure 3: The relic neutralino density, $\Omega h^2$,
 in the $M_2 -
\mu (= - \epsilon)$ plane.}}
\vskip .2in

	The relic abundance of LSP's is controlled by annihilations
until freeze out.  The value of $\Omega h^2$ is roughly proportional
to $1/\langle \sigma v \rangle_{\rm ann}$ and is
determined by solving the Boltzmann
 equation for the LSP number density in an expanding Universe.
 The technique\cite{wso} used is similar to that for computing
 the relic abundance of massive neutrinos\cite{lw}.
 For binos, as was the case for photinos, it is possible
 to adjust the sfermion masses $m_{\tilde f}$ to obtain closure density.
Adjusting the sfermion mixing parameter allows even greater
freedom\cite{fkmos}.
 In figure 3\cite{70}, the relic abundance ($\Omega h^2$)
is shown in the $M_2-\mu$
 plane with $\tan\beta = 2, \mu < 0$, the Higgs
 pseudoscalar mass $m_0 = 50 GeV$,
 $m_t = 100 GeV$ and  $m_{\tilde f} = 200 GeV$.
Binos (which occupy the upper triangular quarter of figure 3
as the LSP), are cosmologically significant in the mass
range $25 - \sim 300$ GeV.  The lower bound coming from the
requirement that for large $\mu$, $M_2 \ga 45$ GeV to aviod a light chargino
(the shaded regions at either large $\mu$ or $M_2$) and the upper
bound coming from the bound on $\Omega h^2$ (heavier binos would require
sfermions with masses $m_{\tilde f} < m_{\tilde B}$).
As annihilations as well as scatterings proceed through sfermion exachange,
detection rates for binos are expected to be somewhat low\cite{fos},
$\la 0.1$/kg/day.
 Clearly the minimal model offers sufficient room
to solve the dark matter problem.
Similar results have been found by other groups\cite{gkt,dn,dvn}.
In figure 3, in the higgsino sector ${\tilde H}_{12}$ marked
off by the dashed line,
 co-annihilations\cite{gs,dn}
 between ${\tilde H}_{(12)}$ and ${\tilde H}_{[12]}$ were not included.
These tend to lower significantly the relic abundance in much of this sector.

\vskip 1in

\centerline{{\tenrm Figure 4: The relic photino density,
  for several values of the stop mass.}}
\vskip .2in

There is also a curious possibility which has been recently
suggested\cite{fmyy}
in which the photino is the LSP and is light and nearly degenerate with a
light stop. For example, it is still experimentally possible that
the lighter stop quark has a mass in the range 20-40 GeV if the
stop mixing angle $\theta_t \simeq 0.98$. At or near this value,
the stop does not couple to the $Z^o$. For a photino with a mass
in the range 16-33 GeV (i.e. nearly degenerate with the stop), the
stop is nearly invisible.
The relic density of the light photinos is acceptable even though
all other sfermion masses may be very high, because the co-annihilation
process\cite{gs} ${\tilde \gamma} + c \rightarrow {\tilde t}$
and ${\tilde t} {\tilde t}^* \rightarrow X$ is efficient if $m_{\tilde t}
- m_{\tilde \gamma} \sim 3 GeV$. The relic density of photinos in this
case\cite{or2} is shown in figure 4.
However, if all other SUSY mass scales are high, this photino is virtually
undectable\cite{or2} although this sector may have consequences
for the top quark branching ratio.

\section{Acknowledgements}
This work was supported in part by  DOE grant DE-FG02-94ER40823.

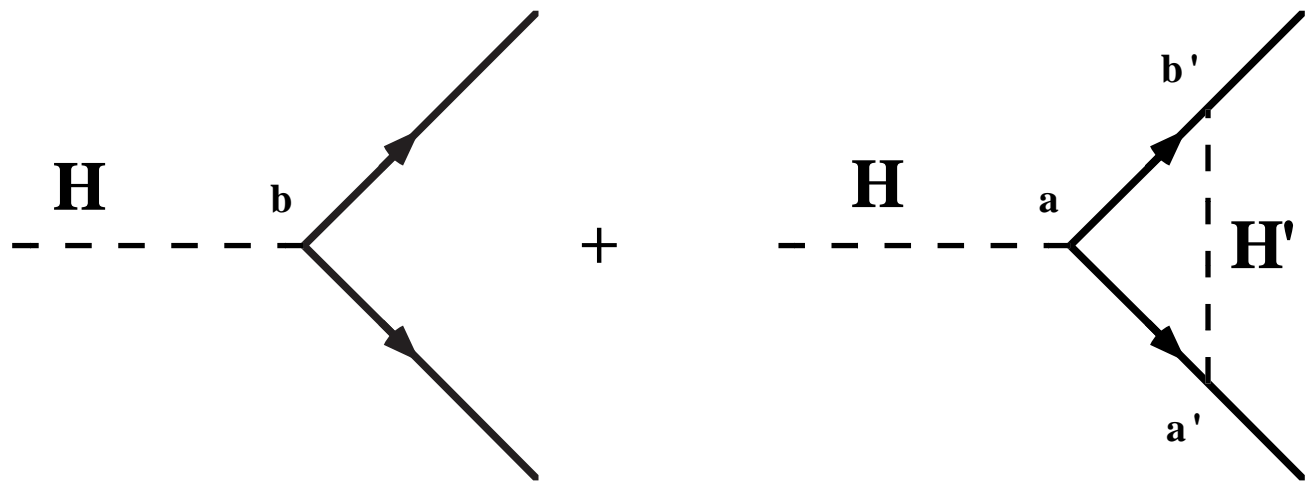
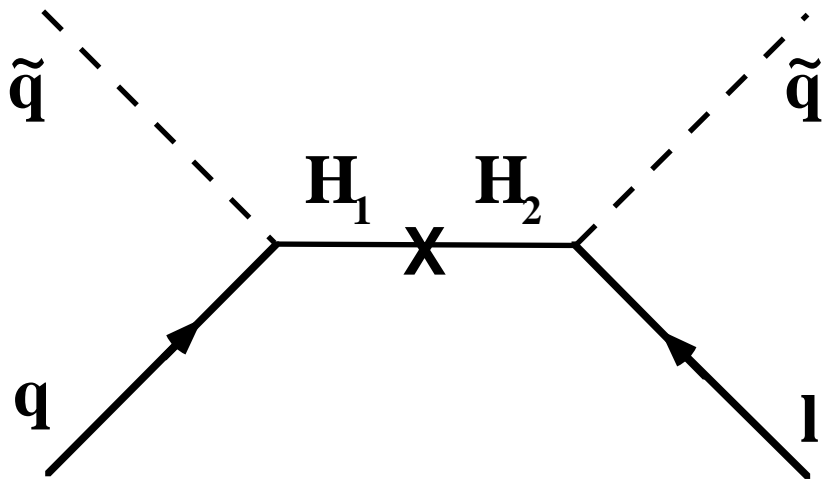
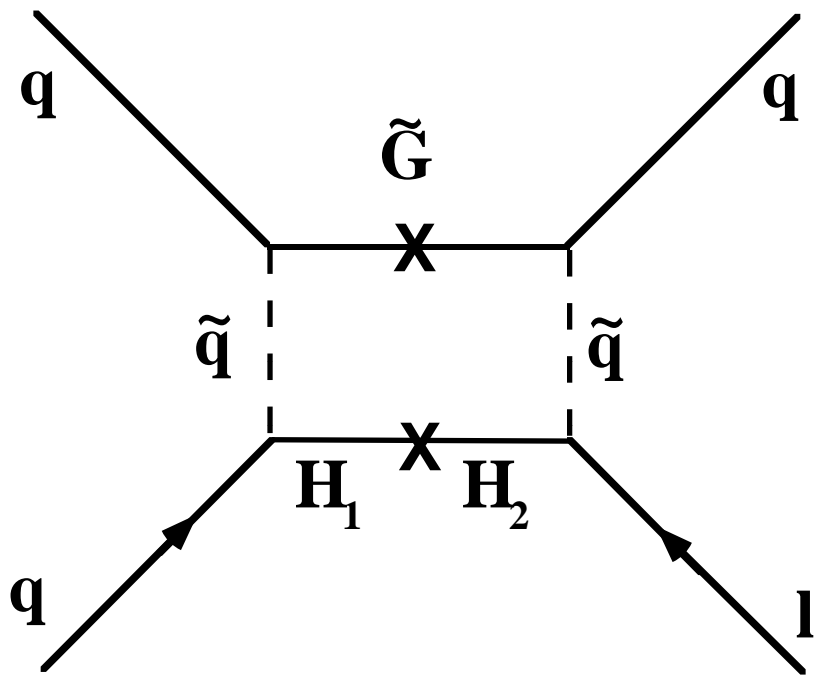
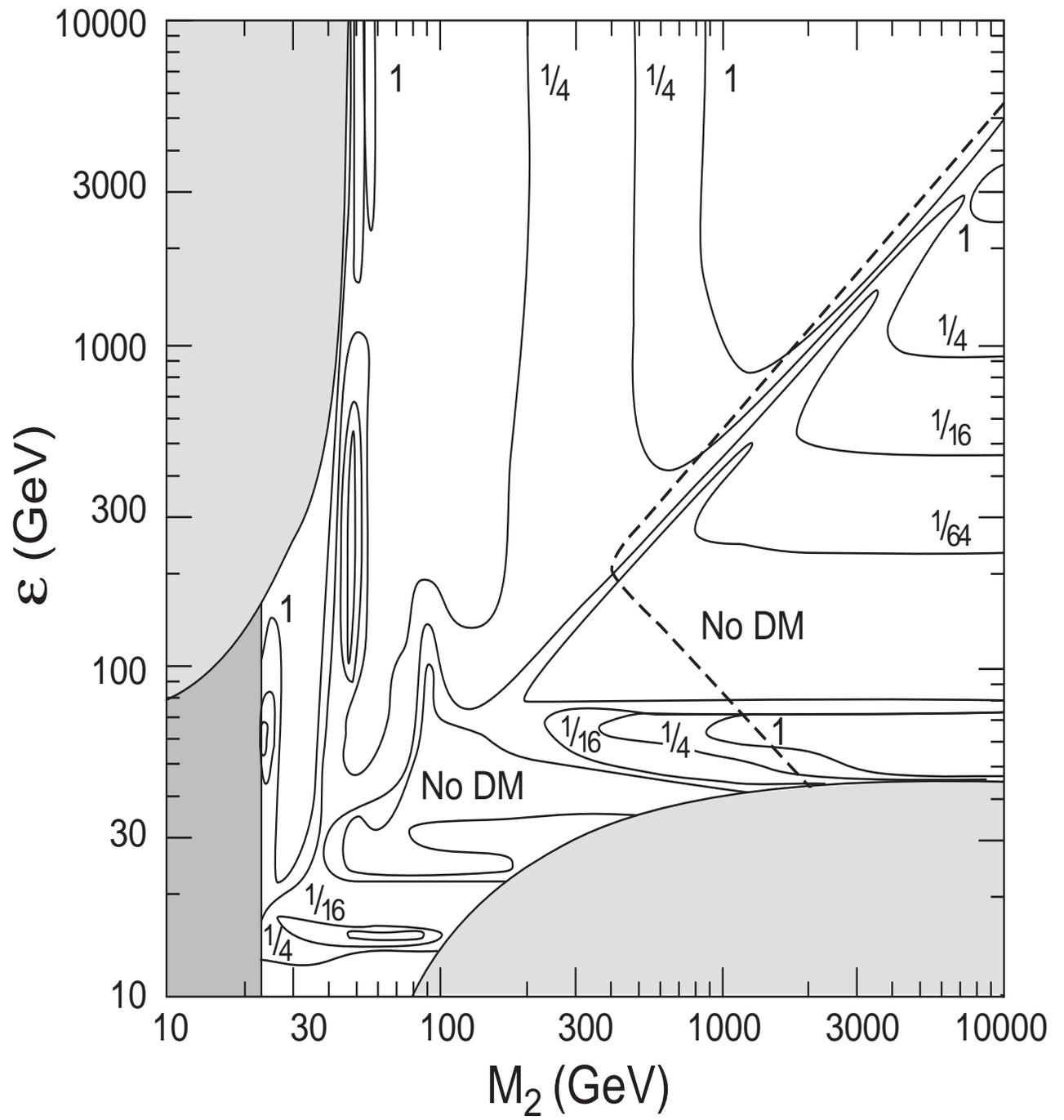
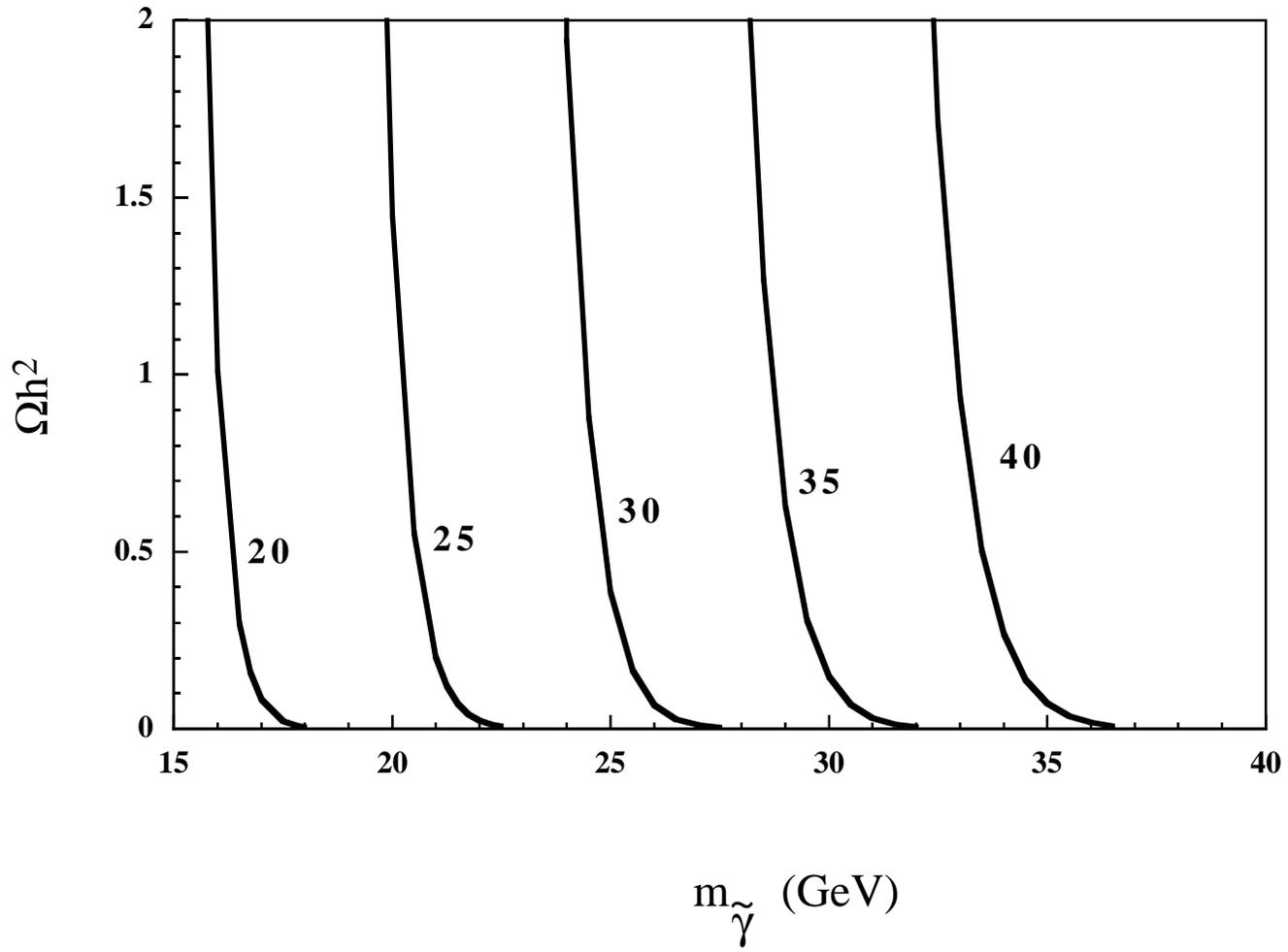
\end{document}